\documentclass[9pt,conference]{IEEEtran}

\usepackage[preprint]{waspaa25}

\usepackage{bm} 
\usepackage{amsmath,amssymb,amsfonts}
\usepackage{algorithmic}
\usepackage{graphicx}
\usepackage{textcomp}
\usepackage{xcolor}
\usepackage{mathtools}
\usepackage{color}
\usepackage{multirow, tabu}
\usepackage{caption, hyperref, enumitem, setspace}
\usepackage{lipsum, url}

\usepackage{booktabs}
\title{TGIF: Talker Group-Informed Familiarization of Target Speaker Extraction}

\name{Tsun-An Hsieh and Minje Kim\thanks{This material is based upon work supported by the National Science Foundation under Grant No. 2512987. We opensource the project at\\\url{https://minjekim.com/research-projects/tgif\#waspaa2025}.}}

\address{University of Illinois Urbana-Champaign, Siebel School of Computing and Data Science,     Urbana, IL 61801, USA}

\begin{document}

\maketitle

\begin{abstract}
State-of-the-art target speaker extraction (TSE) systems are typically designed to generalize to any given mixing environment, necessitating a model with a large enough capacity as a generalist. Personalized speech enhancement could be a specialized solution that adapts to single-user scenarios, but it overlooks the practical need for customization in cases where only a small number of talkers are involved, e.g., TSE for a specific family. 
We address this gap with the proposed concept, talker group-informed familiarization (TGIF) of TSE, where the TSE system specializes in a particular group of users, which is challenging due to the inherent absence of a clean speech target. 
To this end, we employ a knowledge distillation approach, where a group-specific student model learns from the pseudo-clean targets generated by a large teacher model. This tailors the student model to effectively extract the target speaker from the particular talker group while maintaining computational efficiency. 
Experimental results demonstrate that our approach outperforms the baseline generic models by adapting to the unique speech characteristics of a given speaker group. 
Our newly proposed TGIF concept underscores the potential of developing specialized solutions for diverse and real-world applications, such as on-device TSE on a family-owned device. 
\end{abstract}

\section{Introduction}
\label{sec:intro}

Recent advancements in speech font-end applications, such as speech enhancement (SE) \cite{serra2022universal, KoizumiY2023miipher, YangH2024genhancer, LemercierJM2023storm, JukicA2024schrodinger, ScheiblerR2024universe++}, speech separation \cite{lutati2024separate, 
ZhaoS2024mossformer2, SubakanC2023sepformer, ZeghidourN2020wavesplit}, and target speaker extraction (TSE) \cite{Delcroix2018speaker_beam, delcroix_tdSpkBeam, xu2020spex, ge2020spex+} have demonstrated substantial improvements in performance under severe noisy and reverberant acoustic environments. 
However, these advancements often come at the cost of increasing demand for computational resources, which poses significant challenges for on-device applications that require low compute and memory usage.
The growing complexity of these models makes it difficult to deploy them efficiently on edge devices, such as intelligent speakers or smartphones, where such tasks need to run in real-time using only limited computational resources. Hence, research on reducing model complexity for the signal enhancement tasks is one of the essential challenges.  

Among various task-agnostic model compression techniques for audio applications, such as low-bit quantization and pruning \cite{wu2019increasing, KimSW2019icassp, ke2021towards, cohen2023towards, mohamed2024slimmable, miccini2025scalable}, model personalization has emerged as a promising approach, effectively reducing model size while preserving performance.
Model personalization, applied to single-talker speech enhancement tasks, can be seen as a model compression technique, as it reduces the problem space from enhancing \textit{any} speaker's speech to a particular talker. Hence, by relieving the burden of generalization, \textit{personalized speech enhancement} (PSE) could have gained significant complexity reduction while not sacrificing the enhancement quality. 
For instance, the sparse mixture of local experts (MLoE) systems \cite{SivaramanA2021waspaa, ryandhimas2021zero} conduct personalization by assigning mixtures to groups with similar acoustic characteristics and assigning them to local experts. 
Meanwhile, \cite{KimSW2024jasa} uses zero-shot learning combined with knowledge distillation (KD) \cite{HintonG2015arxiv} to personalize SE models without requiring clean speech from the target speaker, where a pretrained large teacher model on a generic dataset guides a smaller student model to adapt effectively to unseen speakers and environments. Finally, it has also been shown that pretraining the SE model from mixtures of noisy speech and additional noise is useful in reducing the amount of personal data required for PSE \cite{SivaramanA2022jstsp}. Note that this PSE concept is different from TSE, which conditions the speech separation model to extract only the target talker, which is sometimes also called personalized speech enhancement \cite{EskimezSE2021personalized}.

However, the successful model compression achieved in PSE has predominantly focused on single-talker contexts, overlooking scenarios involving multiple talkers. Meanwhile, the multi-talker tasks, TSE or speech separation, have not leveraged the fact that the problem space can also be greatly reduced in some common use cases where only a small talker group is involved in the mixture, e.g., recordings from a family-owned smart speaker. 


\begin{figure}[t!]
  \centering
  \centerline{\includegraphics[width=\linewidth]{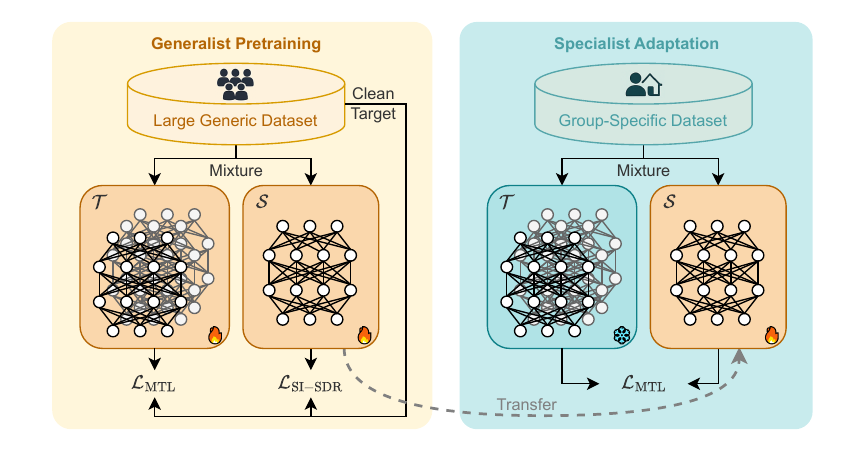}}
  \vspace{-.2cm}
  \caption{An overview of the proposed knowledge distillation process. A generic dataset is used to pretrain the teacher and student generalists. Given pretrained generalists, the student model learns to adapt to a group-specific scenario with the guidance of the pretrained teacher.}
  \label{fig:kd}
\end{figure}

In this paper, we propose a novel talker group-informed familiarization (TGIF) task, where a TSE system gets \textit{familiarized} with a target talker group after the system is deployed to the unseen owner group. In this scenario, talker group-informed TSE aims to optimize model performance for a small group of speakers--an approach that remains underexplored for TSE, yet is practically significant. We see this novel concept as an extension of PSE, where the target adaptation domain grows from a single target talker to a small group of people, e.g., a family. 
To this end, we also propose a novel training framework specifically designed for TGIF of TSE . Our framework employs a two-stage process based on knowledge distillation (KD). In the first stage, a generalist ``teacher" model is pretrained on a large, diverse dataset of noisy reverberant multi-talker speech mixtures without a notion of a speaker group. In the second stage, during test-time adaptation, the teacher model estimates the target speech source from the mixture, which, as ``pseudo" targets, guides the ``student" model's training. Since the test-time adaption lacks access to the clean speech targets of the mixture, the pseudo sources are expected to serve as the proxy target, i.e., the knowledge distilled from the teacher. Meanwhile, since the talker group is a small fixed set of people, we expect that the small student model can still perform well with proper training. Unlike traditional TSE approaches that are trained to generalize to various speaker identities or PSE tailored only for one speaker, our method adapts the student model to the unique acoustic characteristics of a talker group, including reverberation and additional noise, ensuring efficient and personalized performance without the need for clean target data.

We adopt SpEx+ \cite{ge2020spex+}, as a more complex and powerful teacher model to generate high-quality supervision signals. TD-SpeakerBeam \cite{delcroix_tdSpkBeam}, a more lightweight and efficient model, is optimized through KD using SpEx+'s output as pseudo targets. 
The models are evaluated on datasets derived from the DNS Challenge corpus \cite{ReddyC2021dnschallenge}, with experiments across various unseen speakers and acoustic settings to assess performance in realistic, talker group-specific scenarios.
Our results demonstrate that the KD framework can effectively enhance the TSE performance in the proposed TGIF scenarios, and is able to outperform the teacher when there are more interfering speakers, offering a practical solution for applying \textit{familiarized} TSE techniques to a talker group-specific setting.

\section{Talker Group-Informed Familiarization}

\subsection{Problem Definition}

The conventional TSE task begins with a time-domain mixture signal $\bm{x}$ that consists of $K$ speech sources, corrupted by a non-speech source $\bm{n}$ and a room impulse response (RIR) kernel $\bm{h}$ as follows:
\begin{align}
\label{eq:mixing}
    \bm{x}=\bm{h}\ast\bm{s}_{k^*} + \sum_{k=1}^{K-1} \bm{s}_k + \bm{n},
\end{align}
where $\bm{s}_k$ denotes the clean speech source from the $k$-th speaker.
The objective of the TSE model $\mathcal{F}$ is to recover the target source signal $\bm s_{k^*}$ from the mixture, $\bm{s}_{k^*} \approx\hat{\bm{s}}_{k^*} \leftarrow \mathcal{F}(\bm{x}, \bm{e}_{k^*})$, where the enrollment speech of the $k^*$-th speaker, $\bm{e}_{k^*}$, is critical to inform the system of the identity of the target speaker. 

We assume that each speech source $\bm{s}_k$ is spoken by a unique speaker $k$ rather than allowing a mixture of multiple utterances spoken by the same speaker. Hence, $K$ also indicates the number of unique talkers participating in the mixture. In this configuration, it is commonly assumed that the model is tasked with extracting any target speaker within the group, i.e., $k^*\sim\mathbb{G}$, where $\mathbb{G}$ is a set of a very large number of speakers in the conventional definition of TSE.  

Meanwhile, we can also define a \textit{talker group-specific} TSE task, where the input signal $\bm x$ is a mixture of a fixed set of talkers (also corrupted by noise and reverberation), i.e., $k \sim \mathbb{S}^{(l)}$, where $\mathbb{S}^{(l)}$ denotes a talker group with a group index $l \in \{1 \dots L\}$, and $L$ defines the number of groups. Given the definition of the entire talker set $\mathbb{G}$ and the small subset $\mathbb{S}^{(l)}$, let $|\mathbb{G}|$ be the number of total speakers in $\mathbb{G}$. Then, there are $\binom{|\mathbb{G}|}{K}$ combinations of unique speaker mixtures that the TSE model $\mathcal{F}$ should generalize to, while the number of unique speaker mixtures $\binom{|\mathbb{S}^{(l)}|}{K}$ drops exponentially as $|\mathbb{S}^{(l)}|\ll|\mathbb{G}|$. As a result, $\mathcal{F}$ learns to extract the target speaker from the group-specific mixtures more easily.


\subsection{Overview of the Knowledge Distillation Process}

Regardless of its size, architecture, and performance, we call a TSE model a \textit{generalist} if it is trained to extract the target speaker from any given mixture of speakers sampled from $\mathbb{G}$. Otherwise, we call it \textit{specialist}, as it is trained to work only on a small talker group $\mathbb{S}^{(l)}$. Hence, we assume that a small specialist model is a preferred solution in resource-constrained test environments once it works well on the particular speaker group, while it is not required to generalize to other groups. However, it is challenging due to the fact that the speaker group is difficult to be defined ahead of time, leading to a lack of clean speech sources, which is essential for training a model. 

Our approach to the TGIF of TSE employs the KD framework, which was originally proposed for the single-talker PSE problem in \cite{KimSW2021waspaa, KimSW2024jasa}.
As illustrated in Figure \ref{fig:kd}, KD begins with pretraining the generalists. Both the teacher and student models are trained as generalists on a generic, large-scale speaker set $\mathbb{G}$, where the student model is incapable of competing with the teacher's performance.

After a generalist is trained, it is deployed to a test environment. In resource-constrained environments, smaller models are more suitable despite reduced performance, while larger models offer better generalization, but are often impractical due to their cost. In this case, the specialist model could be created by fine-tuning the student generalist using a group-specific dataset. Here, in practice, the group-specific dataset consists of mixture signals only, lacking their dry clean sources. Hence, the KD framework involves the teacher model's output into the training process as the pseudo target. The KD process is a fine-tuning process, as the student model is already pretrained as a generalist, whose model parameters are transferred to initialize the student model before the KD process begins. 
This fine-tuning process adapts the student model to the unique acoustic characteristics and limited set of speakers.

\subsection{Pretraining of Teacher and Student Generalists}
\label{ssec:pretrain}
In the generalist pretraining stage, both the teacher model $\mathcal{T}$ and the student model $\mathcal{S}$ are trained on a generic dataset, which originates from the generic speaker set $\mathbb{G}$.
It is important to synthesize the mixtures with a wide variety of acoustic conditions, including different speakers, noise types, and signal-to-noise ratios (SNRs). In addition, the speakers are randomly selected without considering any specific grouping. Given this condition, the teacher model is trained to extract the target speech from the mixture with reference enrollment: $\bm{s}_{k^*} \approx \hat{\bm{s}}_{k^*}\leftarrow\mathcal{T}(\bm{x}, \bm{e}_{k^*})$. 
The teacher model is a high-capacity network, optimized to produce accurate source estimates $\hat{\bm{s}}_{k^*}$ from $\bm{x}$. 
Concurrently, the student model $\mathcal{S}$ is also trained from the same dataset and procedure, although it tends to underperform the teacher due to its compressed architecture for computational efficiency. 

The training of generalists is straightforward as the loss function can be defined between the predicted source $\hat{\bm{s}}_{k^*}$ and the ground truth source $\bm{s}_{k^*}$. 
The pretraining stage ensures that both models learn general knowledge for handling a variety of acoustic scenarios, providing a robust foundation for the subsequent fine-tuning task on the group-specific data. 
To train the generalists, we minimize the negative scale-invariant signal-to-distortion ratio (SI-SDR) \cite{LeRouxJL2018sisdr} defined between the estimated source and the target source as follows:
\begin{align}\label{eq:kd_tr}
\mathcal{L}_\text{SI-SDR} &\coloneqq
-\text{SI-SDR}\left(\hat{\bm{s}}_{k^*}||\bm{s}_{k^*}\right)\\
\mathcal{L}_{\text{CE}} &\coloneqq -\sum_{k=1}^K y_k \log(\hat{y}_k) \\
\mathcal{L}_{\text{MTL}} &\coloneqq \mathcal{L}_\text{SI-SDR} + \gamma\mathcal{L}_{\text{CE}},
\end{align}
where $\bm y$ and $\hat{\bm y}$ are the $K$-dimensional one-hot representations of target speaker and the prediction from the speaker embeddings, and $\gamma$ serves as a regularization coefficient. Finally, the multi-task learning (MTL) loss $\mathcal{L}_{\text{MTL}}$ is used to train the entire TSE system.
Note that although we include $\mathcal{L}_{\rm CE}$ here, following the best-performing configuration reported in \cite{zmolikova2019speaker}, the student model is trained using only a reconstruction loss $\mathcal{L}_\text{SI-SDR}$. We defer the training details in Sec.~\ref{ssec:setups}.

\subsection{Test-Time Adaptation for Student Specialists}
Test-time adaptation tailors the pretrained student network $\mathcal{S}$ to the specific characteristics of a given talker group.
By introducing a group index $l$, we denote a group of talkers by $\mathbb{S}^{(l)}$.
Accordingly, the $l$-th student model $\mathcal{S}^{(l)}$ is associated with the dataset originates from $\mathbb{S}^{(l)}$, resulting in a TSE model specialized in the acoustic conditions and speaker characteristics of the $l$-th specific group.

Due to the absence of ground-truth clean speech data at test time, the outputs of the generalist teacher model serve as proxy targets. The specialist learns to optimize their performance by minimizing the discrepancy between its outputs $\tilde{\bm s}_{k^*}\leftarrow\mathcal{S}^{(l)}(\bm x, \bm e_{k^*})$ and the corresponding proxy targets $\hat{\bm s}_{k^*}$, thereby adapting effectively to the unique conditions of the talker group $\mathbb{S}^{(l)}$. In summary, the KD loss is defined as follows:
\begin{equation}\label{eq:kd_tt}
\mathcal{L}_\text{MTL}\coloneqq
-\text{SI-SDR}\left(\tilde{\bm{s}}_{k^*}||\hat{\bm{s}}_{k^*}\right)
+\gamma \mathcal{L}_{\text{CE}},
\end{equation}
where the main difference from eq. \eqref{eq:kd_tr} is the teacher's estimate $\hat{\bm s}_{k^*}$ replaces $\bm{s}_{k^*}$ as the learning target for $\mathcal{S}^{(l)}$.
\begin{table*}[thb]
  \centering
  \scriptsize                
  \caption{Mean SI-SDR and SI-SDR improvement. Values in the parenthesis show the improvements from the baseline student generalists.}
  \label{tab:mean_sdr_breakdown}
  \resizebox{0.95\textwidth}{!}{%
    \begin{tabular}{l |rrr  |rrr  |rrr  |rrr  |rrr  |rrr}
      \toprule
      \multicolumn{1}{c}{\multirow{2}{*}{\textbf{Model}}}
        & \multicolumn{3}{c}{\textbf{Overall}} 
        & \multicolumn{3}{c}{\textbf{1 speaker}} 
        & \multicolumn{3}{c}{\textbf{2 speakers}} 
        & \multicolumn{3}{c}{\textbf{3 speakers}} 
        & \multicolumn{3}{c}{\textbf{4 speakers}} 
        & \multicolumn{3}{c}{\textbf{5 speakers}} \\
      \cmidrule(lr){2-4}
      \cmidrule(lr){5-7}
      \cmidrule(lr){8-10}
      \cmidrule(lr){11-13}
      \cmidrule(lr){14-16}
      \cmidrule(lr){17-19}
       & \multicolumn{2}{c}{SI-SDR} & SI-SDRi
       & \multicolumn{2}{c}{SI-SDR} & SI-SDRi
       & \multicolumn{2}{c}{SI-SDR} & SI-SDRi
       & \multicolumn{2}{c}{SI-SDR} & SI-SDRi
       & \multicolumn{2}{c}{SI-SDR} & SI-SDRi
       & \multicolumn{2}{c}{SI-SDR} & SI-SDRi \\
      \midrule

      \scalebox{0.8}{$\mathcal{T}$ }                         
        &  4.66 & & 13.20 
        &  8.60 & & 13.63 
        &  4.32 & & 13.99 
        &  0.66 & & 12.72 
        & –2.03 & & 11.54 
        & –6.18 & &  8.93 \\

      \scalebox{0.8}{$\mathcal{S}_{h=128}$ }               
        &  1.72 & & 10.26 
        &  6.06 & & 11.10 
        &  0.97 & & 10.64 
        & –2.53 & &  9.53 
        & –5.51 & &  8.06 
        & –8.79 & &  6.32 \\

      \scalebox{0.8}{$\mathcal{S}_{h=256}$ }                
        &  2.36 & & 10.90 
        &  6.65 & & 11.68 
        &  1.57 & & 11.24 
        & –1.90 & & 10.16 
        & –4.50 & &  9.07 
        & –8.20 & &  6.91 \\

    \midrule
      \scalebox{0.8}{$\mathcal{S}^{\text{KD}}_{h=128}$}         
        &  2.98 & (+1.26)& 11.52 
        &  6.18 & (+0.12) & 11.22 
        &  2.43 & (+1.46) & 12.10 
        & –0.13 & (+2.40) & 11.92 
        & –2.08 & (+3.43) & 11.49 
        & –5.50 & (+3.29) &  9.61 \\

      \scalebox{0.8}{$\mathcal{S}^{\text{KD}}_{h=256}$}         
        &  3.44 & (+1.08) & 11.97 
        &  6.50 & (-0.15) & 11.54 
        &  3.03 & (+1.46) & 12.70 
        &  0.33 & (+2.23) & 12.38 
        & –1.54 & (+2.96) & 12.03 
        & –4.73 & (+3.47) & 10.38 \\
        
      \scalebox{0.8}{$\mathcal{S}^{\text{KD-Oracle}}_{h=128}$ }  
        &  3.82 & (+2.10) & 12.36 
        &  6.63 & (+0.57) & 11.66 
        &  2.95 & (+1.98) & 12.62 
        &  1.18 & (+3.71) & 13.23 
        & –0.27 & (+5.24) & 13.30 
        & –2.22 & (+6.57) & 12.89 \\

      \scalebox{0.8}{$\mathcal{S}^{\text{KD-Oracle}}_{h=256}$}
        &  4.42 & (+2.06) & 12.96
        &  7.05 & (+0.40) & 12.08 
        &  3.74 & (+2.17) & 13.41 
        &  1.88 & (+3.78) & 13.93 
        &  0.57 & (+5.07) & 14.14 
        & –1.78 & (+6.42) & 13.34 \\

      \bottomrule
    \end{tabular}%
  }
\end{table*}

\section{Experiments}
\subsection{TGIF Dataset}
We create our TGIF dataset based on the ICASSP 2021 Deep Noise Suppression (DNS) Challenge \cite{ReddyC2021dnschallenge}. For generalist training, the dataset is talker-group agnostic, meaning that each mixture can contain speech from \textbf{any speaker} in $\mathbb{G}$. During test time, a talker group is postulated to simulate the real-world family environment, where all speakers in the mixture are drawn exclusively from \textbf{the same talker group $\mathbb{S}^{(l)}$}. 
The noise and reverberation conditions differ between generalist pretraining and test-time adaptation. Pretraining covers a broad range of noise types and room acoustics, while test-time adaptation is limited to household noises and a narrower range of room sizes. 
  
\subsubsection{Generic Dataset for Generalist Pretraining}
We partially follow the data preparation recipe of the DNS Challenge, with modifications to incorporate multi-speaker scenarios as shown in eq.~\eqref{eq:mixing}. Clean speech segments are sourced from LibriVox~\cite{kearns2014librivox}, and the noise clips are selected from AudioSet \cite{GemmekeJ2017audioset} and Freesound \cite{fonseca2017freesound} as in the DNS Challenge.

To simulate reverberation, we sample RIRs from the OpenSLR28 dataset \cite{ko2017study}, which provides diverse acoustic profiles in room sizes and reverberation times. We apply the sampled RIR only to the target speaker's utterance, 80\% of the time, while the other sources remain dry. 
Each audio mixture is 10 seconds long and contains between one and five speakers, with signal-to-interference ratios (SIRs), defined between the reverberant target utterance and the \textit{sum} of other speech utterances, uniformly sampled from the range of $[-5, 25]$ dB. While we ensure the overall SIR of the speech mixture, each interfering speech utterance's loudness is also randomized using the same range. 
Background noise is then added to these reverberant speech mixtures, controlled by a signal-to-noise ratio (SNR) drawn from the same $[-5, 25]$ dB range, measured specifically between the reverberant target speech and the noise. 
After all, each mixture includes up to one noise source and up to four interfering speakers, possibly overlapped. In total, we generate $1,000$ hours of training data and 50 hours for validation.
For each training example, a 3-second enrollment is selected from a different utterance spoken by the same target speaker of the given mixture. As a result, both the mixture and enrollment signals are three seconds long as inputs to the TSE model.

\subsubsection{TGIF Dataset for Test-time Adaptation}\label{sec:tgif_data}
For testing, we construct 20 talker group-specific datasets using a separate set of clean speech, noise, and RIRs, sourced from different datasets: clean speech from the Voice Cloning Toolkit (VCTK) corpus \cite{veaux2016superseded}, noise from DEMAND \cite{ThiemannJ2013DEMAND}, and RIRs from OpenSLR26 \cite{ko2017study}.


We assign up to five \textit{fixed} members to each test group that are mixed similarly to the generic dataset along with a randomly selected household noise source and the RIRs of a fixed room to create 10-second mixtures. We assume that there are clean enrollment segments for the individual members. 
For each group $\mathbb{S}^{(l)}$, the target speaker in every mixture is randomly selected from the group's members, and the number of unique speakers $K$ ranges from one to $|\mathbb{S}^{(l)}|$.
For the talker group-specific dataset, we maintain the same SIR range used in the generic dataset, but adjust the SNR range to $[-15, 15]$~dB. This modification creates a noisier acoustic environment for us to evaluate both the generalization capabilities of the baseline models and the effectiveness of KD-based adaptation.
In total, we synthesize 50 hours of mixture data: 20 hours for KD adaptation, 10 hours for validation to prevent overfitting, and 20 hours for evaluating the student specialists. Unlike generalist pretraining, test-time adaptation is tailored to household acoustic environments, using only noise types such as dishwashing, kitchen activity, and living room ambiance, within small-room reverberation settings.

\subsection{Baseline Models Configurations}
Our student models adopt the time-domain (TD) SpeakerBeam \cite{zmolikova2019speaker} architecture. TD-SpeakerBeam works in time domain and includes a convolutional encoder-masker-decoder structure. Speaker embeddings are fused at the 7th layer of the masker using a multiplicative conditioning mechanism.

We train two student model variants with reduced hidden dimensionality to evaluate the impact of our adaptation strategy. While the original TD-SpeakerBeam uses a hidden size of 512, our student models adopt smaller configurations with hidden size of 256 and 128. This setup allows us to examine whether KD can effectively improve test-time performance even in more compact models.

We employ SpEx+ \cite{ge2020spex+} as our teacher model. SpEx+ is a also a fully TD architecture that improves upon its predecessor \cite{xu2020spex}. Different from TD-SpeakerBeam, SpEx+ consists of two parallel encoders, for the mixture and the enrollment utterance, respectively, that share weights to ensure consistency in the learned representations in early layers. These encoders are followed by a temporal convolutional network (TCN)-based separator, which predicts a mask conditioned on the target speaker's identity. The speaker embedding, extracted from the enrollment encoder, uses a ResNet-based \cite{HeK2016cvpr} speaker encoder, and is injected into the separator via affine transformations at multiple layers. Finally, a decoder reconstructs the waveform of the target speaker from the masked latent features.

\subsection{Pretraining \& KD-Based Adaptation Setup}
\label{ssec:setups}
We pretrain both the teacher and student generalists using the setup described in Section~\ref{ssec:pretrain}. Following \cite{zmolikova2019speaker, ge2020spex+}, we set the regularization coefficient $\gamma$ to 0 and 0.5 respectively for student and teacher models. The generalists are trained for up to $1,000$ epochs with batch sizes of 8 and 16 for $\mathcal{T}$ and $\mathcal{S}$, respectively, using the Adam \cite{KingmaD2015adam} optimizer with an initial learning rate of 0.001. A patience-based scheduler halves the learning rate if no improvement is observed for 20 validation epochs. Early stopping is employed with a patience of 120 epochs based on validation loss. The mixture and enrollment segments are cropped to 3 seconds and are resampled to 16~kHz.

For test-time adaptation, we distill the pseudo-target from the SpEx+ teacher model and use it to fine-tune the pretrained TD-SpeakerBeam student generalists. The adaptation is performed separately for each talker group from the TGIF testing set (Sec. \ref{sec:tgif_data}). Each mixture and enrollment pair is 10 seconds long, and both are sampled at 16~kHz. The student model is initialized from the best checkpoint obtained during generalist pretraining and fine-tuned for up to 120 epochs using the Adam optimizer with a learning rate of $10^{-5}$. Early stopping is not applied for adaptation.    

To evaluate the effectiveness of adaptation under both ideal and more realistic conditions where clean references may not be available, the adaptation process considers both oracle and teacher-forcing adaptation scenarios. In the oracle setup, the student specialists are fine-tuned using the ground truth clean target speech as the supervision signal. In the teacher-forcing setup, which follows a KD paradigm, the model instead uses the output of the SpEx+ teacher as a pseudo target.

\begin{figure}[t]
  \centering
  \centerline{\includegraphics[width=0.95\linewidth]{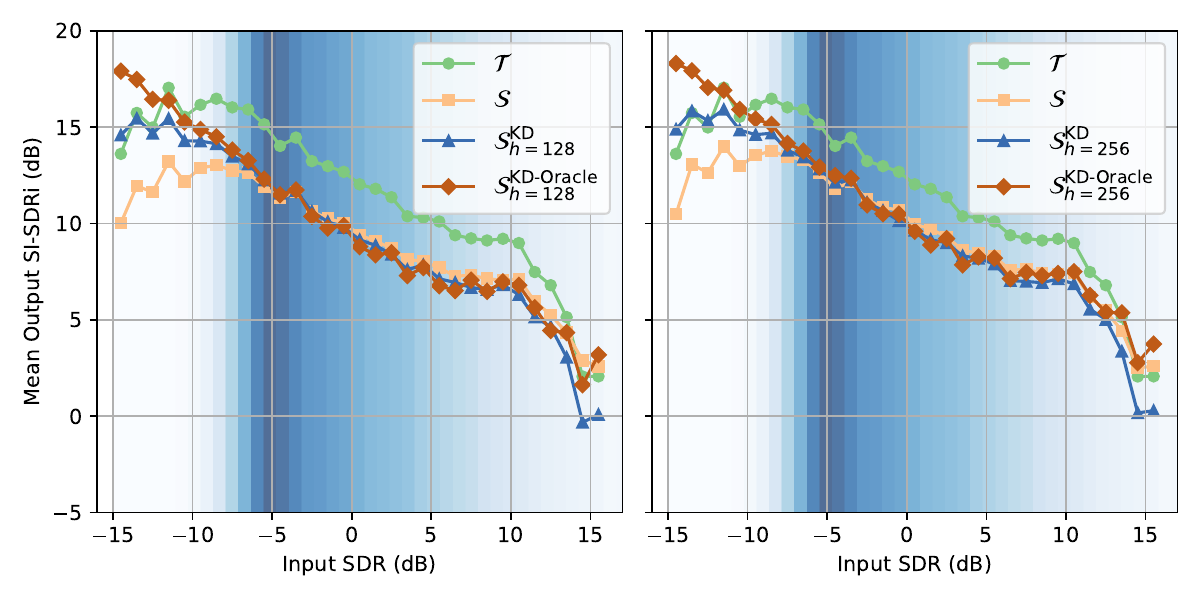}}
  \vspace{-.2cm}
  \caption{SI-SDRi trends across input SDR.}
  \label{fig:sdr_vs_si_sdr}
\end{figure}
\section{Results and Analysis}

Table~\ref{tab:mean_sdr_breakdown} and Fig.~\ref{fig:sdr_vs_si_sdr} summarize the extraction performance for the teacher model ($\mathcal{T}$), the generalist student ($\mathcal{S}$), the student specialist fine-tuned via KD ($\mathcal{S}^{\text{KD}}$), and the oracle KD variant ($\mathcal{S}^{\text{KD-Oracle}}$). The results are analyzed across different numbers of interfering speakers and input SDR ranges.

\subsection{Overall Performance}

As shown in Table~\ref{tab:mean_sdr_breakdown}, overall, $\mathcal{T}$ consistently achieves better mean SI-SDR and SI-SDRi performance  (4.66~dB and 13.20~dB, respectively) than the student models. The generalist students without fine-tuning perform substantially worse, with a notable drop of 2.94~dB particularly for the smaller model. The KD student models improve the generalist baseline, by closely approaching the teacher's performance with the gaps of 1.68~dB and 1.22~dB for $\mathcal{S}^{\text{KD}}_{h=128}$ and $\mathcal{S}^{\text{KD}}_{h=256}$, respectively, which is a 1.26 dB improvement from the small generalist student.  
In contrast, the larger oracle KD variant achieves further gains, reducing the SI-SDR gap relative to the teacher to 0.24~dB.

\subsection{Impact of Number of Speakers}

As the number of speakers increases from two to five, all models’ performance declines, maintaining similar gaps between $\mathcal{T}$ and the student baselines. However, the proposed TGIF framework shows more significant merit in the harsher input conditions. For example, in the single‐speaker case, the fine‐tuned student specialists ($\mathcal{S}^{\text{KD}}$) are on par with their generalist counterparts.
In contrast, in the two‐ to five‐speaker scenarios, every $\mathcal{S}^{\text{KD}}$ model outperforms its generalist version: $\mathcal{S}^{\text{KD}}_{h=128}$ outperforms $\mathcal{S}_{h=128}$ by 1.46~dB, 2.40~dB, 3.43~dB, and 3.29~dB for two, three, four, and five speakers, respectively, showing that KD-based TGIF yields increasingly larger gains in more complex acoustic environments. Interestingly, the TGIF student models even surpass $\mathcal{T}$ in the four‐ and five‐speaker conditions. This suggests that the TGIF configuration reduces the speaker encoder's burden of discriminating thousands of diverse speakers to only a few group members, leading to improved contrasts between the speakers performance and highlighting its effectiveness for model compression.

Comparing oracle and non‐oracle setups reveals that the benefit of teacher quality grows with task difficulty. The SI-SDR gaps between $\mathcal{S}^{\text{KD-Oracle}}_{h=128}$ and $\mathcal{S}^{\text{KD}}_{h=128}$ widen from 0.45~dB (one speaker) to 0.52~dB, 1.31~dB, 1.81~dB, and 3.28~dB as we move to five speakers. This trend indicates that high‐quality teacher outputs become increasingly critical under heavier interference.

\subsection{Performance Trends Across Input SDR}

Fig.~\ref{fig:sdr_vs_si_sdr} visualizes model behaviors across different input signal-to-distortion ratio (SDR) values. We define SDR as the overall input signal's quality, where the signal is defined as the dry clean target utterance, while the distortion is the difference between the reverberant input mixture and the target. The blue heatmap indicates the density of training samples across different input SDR intervals. Note that the overall SDR metric can be lower than the SIR or SNR range $[-5, 25]$ dB used for data preparation, considering additive noise and reverberation. The upper and lower plots show the mean SI-SDRi scores of various models as a function of input SDR.

At very low input SDRs, around -14~dB, where training samples are sparse, $\mathcal{S}^{\text{KD-Oracle}}$ achieves the highest SI-SDR improvement, followed by $\mathcal{S}^{\text{KD}}$ and $\mathcal{T}$, while $\mathcal{S}$ falls significantly behind. Notably, $\mathcal{S}^{\text{KD}}$ outperforms $\mathcal{T}$ in this region, despite relying on the teacher's guidance during KD. This trend highlights that familiarization, whether via KD or oracle finetuning, is crucial for TSE when the speech sources are highly overlapped.
Starting from approximately -4~dB, the effectiveness of KD diminishes and fails to surpass the student generalist $\mathcal{S}$, even with the oracle KD setup.
In the high input SDR region ($>$13~dB), where mixtures are relatively clean, all models converge to modest SI-SDR gains (0–5~dB), leading KD versions to perform the worst. 

In summary, the proposed TGIF method proves especially effective in low-quality and out-of-domain regions, where training samples are sparse and generalist models struggle to generalize. 

\section{Conclusion}

This work introduces a new TGIF task, extending the single-speaker personalized speech enhancement systems to TSE for a small group of people, such as a family. By adapting a lightweight student model through KD using a teacher model’s pseudo-labels, the specialists significantly improve the performance, especially under unseen and challenging conditions. Overall, the specialists surpass their generalist counterparts by over 1~dB. Under severe conditions, e.g., multiple interfering speakers or low input SDR, the specialists even surpass the teacher, demonstrating their ability to adaptively discriminating group members for robust TSE.



\clearpage
\bibliographystyle{IEEEtran}
\bibliography{minje}







\end{document}